\pdfoutput=1

\documentclass[11pt]{asaproc}%
\usepackage{natbib}
\usepackage{graphicx}
\usepackage{verbatim}
\usepackage{amsmath}
\usepackage{amsfonts}
\usepackage{amssymb}
\usepackage{url}%
\providecommand{\U}[1]{\protect\rule{.1in}{.1in}}
\begin{document}

\title{Bayesian Tracking of Emerging Epidemics Using Ensemble Optimal Statistical
Interpolation (EnOSI)}
\author{Ashok Krishnamurthy\thanks{ashok.krishnamurthy@ucdenver.edu}
\and Loren Cobb${}^{*}$
\and Jan Mandel${}^{*}$
\and Jonathan Beezley${}^{*}$}
\maketitle

\begin{abstract}
We explore the use of the optimal statistical interpolation (OSI) data
assimilation method for the statistical tracking of emerging epidemics and to
study the spatial dynamics of a disease. The epidemic models that we used for
this study are spatial variants of the common susceptible-infectious-removed
(S-I-R) compartmental model of epidemiology. The spatial S-I-R epidemic model
is illustrated by application to simulated spatial dynamic epidemic data from
the historic "Black Death" plague of 14th century Europe. Bayesian statistical
tracking of emerging epidemic diseases using the OSI as it unfolds is
illustrated for a simulated epidemic wave originating in Santa Fe, New Mexico.


\end{abstract}

\section{Introduction\label{intro}}

Mathematical models have been used since 1927 to describe the rise and fall of
infectious disease epidemics \citep{Diekmann-2000-MEI, Castillo-2002-MAE1,
Castillo-2002-MAE2, Ma-2009-MUI}. A\ majority of the models are based on the
three-compartment nonlinear Susceptible-Infected-Removed (S-I-R) model
developed by \cite{Kermack-1927-ACM}. A person occupies the \emph{susceptible}%
\ or \emph{infectious}\ compartments if he or she can contract or transmit the
disease, respectively. The \emph{removed}\ compartment includes those who have
died, have been quarantined, or have recovered from the disease and become
immune. The state variables are the number of susceptible ($S$), the
infectious ($I$), and the removed ($R$) in a closed population. S-I-R models
often perform surprisingly well in modeling the temporal evolution of
real-world epidemics, and their spatial variants can produce a traveling-wave
spatial dynamics similar to that displayed by some epidemics. Traveling waves
are solutions to the spatial models such that the distribution of infected at
time ($t+1$) is approximately a translation of the distribution at time $t$.

Tracking and forecasting the full spatio-temporal evolution of a new epidemic
is notoriously difficult. Often the model itself is incorrect in unknown ways,
observational data may be affected by many sources of error, and new data
arrives on an irregular schedule. This is a well-known problem in a variety of
empirical areas of high importance, such as storm and wildfire forecasting.
The general category of tracking techniques that incorporate error-prone data
as they arrive by sequential statistical estimation is known as \textit{data
assimilation} \citep{Kalnay-2003-AMD}. Use of the statistical methods of data
assimilation can increase the accuracy and reliability of epidemic tracking by
incorporating data as it arrives, with weighting factors that reflect the
observed reliability of the observations. A few applications of data
assimilation in epidemiology already exist \citep{Kalivianakis-1994-RSV,
Cazelles-1997-UKF, Costa-2005-MPA, Bettencourt-2007-TRT, Bettencourt-2008-RTB,
Jegat-2008-EDA, Rhodes-2009-VDA, Dukic-2009-TFE, Mandel-2010-DDC}.

The aim of this paper is to study the use of a spatial variant of the S-I-R
model to track newly emerging epidemics using a data assimilation technique
called optimal statistical interpolation (OSI). This is already a popular data
assimilation method in the literature of meteorology and oceanography. When
coupled with a spatial dynamic model, the OSI method can be used to forecast
the spatio-temporal evolution of an epidemic, and to adjust those forecasts
appropriately as sparse and error-prone data arrives.

This paper is organized as follows. In Section \ref{epi}, we present a
stochastic spatial epidemic model and use it to reproduce the spatio-temporal
disease spread map of the 14th century Black Death. In Section \ref{track}, we
illustrate the Bayesian tracking of emerging epidemics using OSI with a
simulated epidemic wave originating in Santa Fe, New Mexico. In Section
\ref{results}, we provide computational results for the stochastic spatial
epidemic model with epidemic tracking method presented in Section \ref{track}.
Finally, in Section \ref{conclusion}, we provide some concluding remarks and
future directions.

\section{A Stochastic Spatial Epidemic Model\label{epi}}

\subsection{Epidemic Dynamics}

For this study we use a discretized stochastic version of the
\cite{Hoppensteadt-1975-PDE} spatial S-I-R epidemic model. As with almost all
spatial epidemic models since \cite{Bailey-1957-MTE, Bailey-1967-SSE,
Kendall-1965-MMS}, we assume that individuals are continuously distributed on
a spatial domain. This model uses three variables to define the state of the
epidemic at each $(x,y)$ coordinate:\bigskip

$\qquad S(x,y,t)=$ density (per unit area) of the susceptible population,

$\qquad I(x,y,t)=$ density of the infected population, and

$\qquad R(x,y,t)=$ density of the removed population.\bigskip

Thus, each of these variables is a scalar field that evolves with time.

In continuous time the epidemic dynamics are defined by a system of three
partial differential equations for the state variables. There are no vital
dynamics in this model, meaning that there are no new births or non-disease
related deaths in any of the three compartments. Following
\cite{Hoppensteadt-1975-PDE}, we assume that the rate of new infections at
location $(x,y)$ depends on the density of infection at that point, and in
nearby locations that have been weighted with a kernel function that drops off
exponentially with Euclidean distance. The effective (i.e. weighted) density
of infection at $(x,y)$ is given by
\begin{align*}
J(x,y,t)  &  =\int\int I(x-\phi,y-\theta)K(\phi,\theta)d\phi d\theta,\\
K(\phi,\theta)  &  \propto\exp\left(  -\alpha\sqrt{\phi^{2}+\theta^{2}%
}\right)  ,
\end{align*}
where the integral is taken over the entire surface area under study, and the
proportionality constant is given by the condition that $\int\int
K(\phi,\theta)d\phi d\theta=1$. Then, the three partial differential equations
for the \textit{deterministic } evolution of the spatial epidemic are:%
\begin{align*}
\partial_{t}S  &  =-\beta SJ,\\
\partial_{t}I  &  =\beta SJ-\gamma I,\\
\partial_{t}R  &  =\gamma I.
\end{align*}

In this model, $\beta$ is the rate of infection from infected to susceptibles,
given homogeneous mixing, $\alpha$ is an intensity measure of infectiousness
of the disease, given by the product of mixing rate and the infection rate and
$\gamma$ is the rate of removal of infected persons through death, recovery
with immunity, and quarantine. To make this model stochastic, we assume that
the quantities $\beta SJ$ and $\gamma I$ are the intensities of two
independent Poisson processes. For simulations in discrete time and space, we
use the following approximation: The number of newly infected and newly
removed persons over the time interval $(t,t+\Delta t)$, within a box centered
at position $(x,y)$ with $\Delta x\Delta y$ units of area, are given
by\bigskip

\qquad Number newly infected $\sim Poisson(\beta S(x,y,t)J(x,y,t)\Delta
x\Delta y\Delta t)$,

\qquad Number newly removed $\sim Poisson(\gamma I(x,y,t)\Delta x\Delta
y\Delta t).$\bigskip

Thus a susceptible individual, at a particular location, may become infected
when he/she comes in contact with an infected individual from within a
neighboring area, with a monotonically decreasing weighting function that
declines exponentially with distance. If this contact causes sufficient
secondary infections then a new epidemic focus will develop at that new
location. The simulation evolves on a two-dimensional discretized spatial
domain with a total of $n\times m$ grid cells.

\subsection{Example: The Black Death in Europe}

The "Black Death" bubonic plague epidemic that hit Europe in 1347 killed
somewhere between 30\% and 60\% of the population of Europe over the course of
about four years. The virulence of the disease back then was severe, which
explains the extremely high number of deaths. Plague recurred in various
regions of Europe for another 300 years before gradually withdrawing from
Europe. There have been several attempts to reconstruct the movement of the
wavefront of the epidemic \citep{Langer-1964-TBD, Noble-1974-GTD,
Christakos-2005-NST, Christakos-2007-RRS, Gaudart-2010-DAS}, as it swept
across the continent; one such attempt using our spatial S-I-R model is shown
in Figure 1b; others are similar. The disease arrived from Asia into Europe by
trading ships, appearing first in Constantinople (modern Istanbul, Turkey) in
1347. From there it was carried by ship to Italy, France, Spain, and Croatia.
Once ashore it moved inland (mainly in pneumonic form) at a speed that has
been estimated at between 100 and 400 miles per year.

If it is to have credibility, a spatial epidemic model should be able to
reproduce the principal historical features of the Black Death: its movement
into the interior of Europe from the coastal cities, especially Marseilles,
and movement up the island of Britain after its arrival in Bristol and London.
Figure 1b shows the population density of infected people, using modern
population densities in place of the unknown medieval population pattern, at
roughly the beginning of 1350.

\begin{figure}
[ptb]
\begin{center}
\fbox{\includegraphics[
natheight=7.735700in,
natwidth=16.611300in,
height=2.5036in,
width=5.3437in
]%
{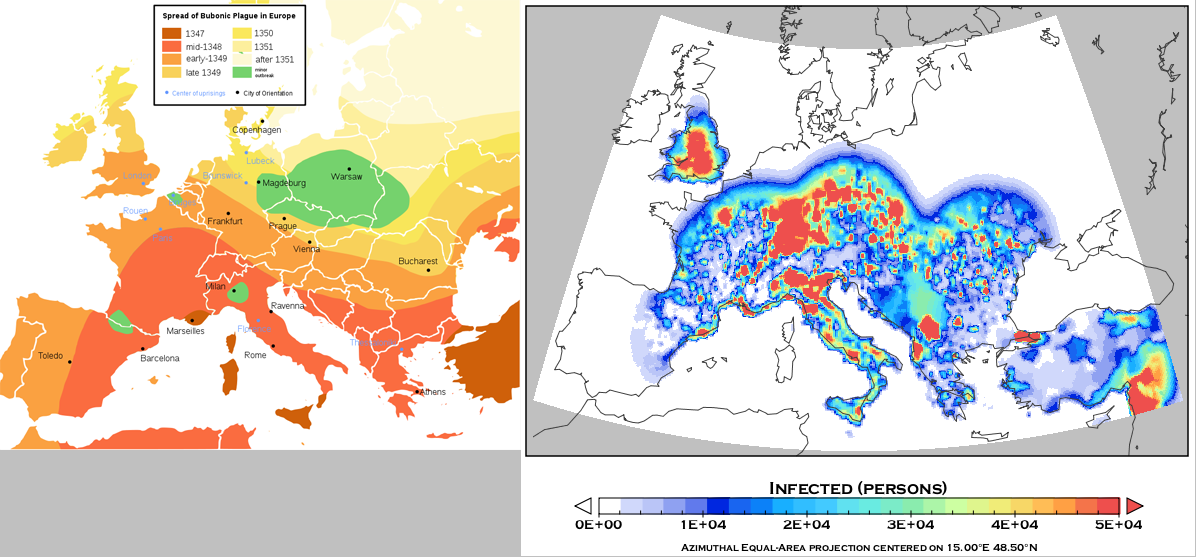}%
}\caption{Figure 1a (left) Historical reconstruction. (Reproduced from
\cite{Zenner-2008-DPN} under Creative Commons license.) 1b (right) Simulated
state of the epidemic in 1350.}%
\end{center}
\end{figure}

The R statistical computing language \citep{R-2010-LES}, freely available from
\url{www.cran.r-project.org}%
, was used to carry out the simulations for the spatial spread of the epidemic.
Modern day population density data were downloaded as GPW
(Gridded Population of the World) data files from the Center for International
Earth Science Information Network at the Columbia University and converted to
the array-oriented Network Common Data Form (NetCDF) format. These datasets
were then loaded into R using the built-in package \emph{ncdf}.

\subsection{Comparison to Historical Data}

In historical reality, the zone from Romania to Poland to Russia suffered only
lightly from the first wave of the Black Death. It is quite plain that the
S-I-R simulation does not conform to the historical reconstruction in that
zone. The explanation may lie with the very low population density and
geographic mobility in Eastern Europe in the 14th century. A smaller
discrepancy occurs in the city of Milan, Italy, which brutally stopped the
spread of the plague by boarding up infected people in their homes. Despite
differences in population distribution (France was then much more populous
than Germany, for example), we believe that the simulation performs reasonably
well in all areas except Eastern Europe.

\section{Statistical Tracking Using Data Assimilation Techniques\label{track}}

We use data assimilation for the statistical tracking of emerging epidemics as
they are unfolding. This involves two basic components: a dynamic model to
forecast the state of the epidemic between arrivals of new data, and
observations that are used to update an ensemble of state estimates. Data
assimilation requires estimating the uncertainty both for model and
observations forecasts. Our goal in this paper is to incorporate sparse and
noisy observational epidemic data over space and time into a dynamic
statistical model so as to produce an optimal Bayesian estimate of the current
state of the infected population, and to forecast the progress of the real epidemic.

\subsection{The Kalman Filter (KF)\label{kf}}

The Kalman Filter (KF) was first presented by \cite{Kalman-1960-NAL} and
\cite{Kalman-1961-NRL} as a method for tracking the state of a linear dynamic system
perturbed by Gaussian white noise. In mathematical terms, this means that the
errors are drawn from a zero-mean distribution with diagonal covariance matrix.

The full state of a discretized spatial epidemic model is a grid of $n\times
m$ cells, each of which contains a characterization of the population
currently within the limits of the cell. To apply the Kalman filtering method,
we represent the $n\times m$ values of the \textit{Infected} variable on this
grid\ as a single long vector $x$\ with $p=n\times m$ elements, which for the
purposes of data assimilation is the dimensionality of the state space. If we
could observe this state vector without error, our observations would be
another vector $y$ that satisfies $y=Hx$. Now consider the situation in which
we have a forecast of the current state, $x^{f}$, and a newly arrived vector
of noisy observations $y=Hx+\varepsilon$, where $\varepsilon\sim N\left(
0,R\right)  $. We need to update the forecast by optimally assimilating these
new observations. The result will be called the \textit{analysis} estimate of
state, $x^{a}$. The superscripts $f$ and $a$ are used to denote the forecast
(prior) and analysis (posterior) estimate of the current state, respectively.

In the classical Kalman filter, the underlying dynamics are assumed to be
linear, e.g.%
\begin{align*}
x_{t}  &  =F_{t-1}x_{t-1}+u_{t-1},\\
u_{t}  &  \sim N\left(  0,\Sigma\right)  ,
\end{align*}
and the analysis estimate of the state vector is calculated from%
\begin{align*}
x_{t}^{f}  &  =F_{t-1}x_{t-1}^{f}\\
x_{t}^{a}  &  =x_{t}^{f}+K_{t}(y_{t}-Hx_{t}^{f}),\\
Q_{t}^{a}  &  =(I-K_{t}H)Q_{t}^{f},
\end{align*}
where%
\begin{align*}
K_{t}  &  =Q_{t}^{f}H^{T}(HQ_{t}^{f}H^{T}+R)^{-1},\text{ and}\\
Q_{t}^{f}  &  =F_{t-1}Q_{t-1}^{f}F_{t-1}^{\prime}+\Sigma.
\end{align*}
Here $K_{t}$\ is the Kalman gain matrix at time $t$, and $Q_{t}^{f}$ is the
covariance matrix for the forecast state vector.

The extended Kalman filter (EKF) \citep{Julier-1995-ANA} was an early attempt
to adapt the basic KF equations for nonlinear problems, through linearization.
However, the EKF has it's own disadvantages: if the model is strongly
nonlinear at the time step of interest, linearization errors can turn out to
be non-negligible, which leads to filter divergence \citep{Evensen-1992-UEK}.
The EKF is not suitable for high-dimensional 2D and 3D data assimilation
problems. Other commonly used Bayesian tracking techniques for nonlinear
problems include the unscented Kalman filter (UKF) and the particle filter
(PF) \citep{Gordon-1993-NAN}. Particle filtering is a versatile Monte Carlo
technique that can handle nonlinearities in the model and represents the
Bayesian posterior probability density function by a set of samples drawn at
random with associated weights.

\subsection{Ensemble Kalman Filter (EnKF)\label{enkf}}

The KF algorithm described above is based on assimilating only one initial
state ignoring the uncertainties in the model. In the following, we account
for this state-dependent uncertainty by taking an ensemble approach to data
assimilation. To see the problem, consider a 2D simulation of a scalar field
that has been discretized on a $10^{3}\times10^{3}$ grid. The state vector of
this system has $10^{6}$ elements, and its covariance matrix has $10^{6}%
\times10^{6}=10^{12}$ elements, requiring eight terabytes just to store. The
EnKF solves this storage-and-retrieval problem by (in effect) calculating the
covariances from the members of the ensemble as they are needed. The result is
in an elegant Bayesian update algorithm with dramatically improved efficiency
and storage requirements \cite{Mandel-2010-DDC}.

The ensemble Kalman filter (EnKF) was introduced by 
\cite{Evensen-1994-SDA}, modified to provide correct covariance by 
\cite{Burgers-1998-ASE}, and improved by 
\cite{Houtekamer-1998-DAE}. The EnKF is a popular sequential Bayesian data
assimilation technique that uses a collection of almost-independent
simulations (known as an ensemble) to solve the covariance problem of
Kalman filtering for systems with very high-dimensional state vectors. It does
this using a two-step process: estimate of the covariance matrix, followed by
an ensemble update. The covariance of a single state estimate in the KF is
replaced by the sample covariance computed from the ensemble members. This
sample covariance of ensemble forecasts is then used to calculate the Kalman
gain matrix. There are two basic approaches to the EnKF update: stochastically
perturbed observations (Monte Carlo), and \textquotedblleft
square-root\textquotedblright\ filters (deterministic). Both approaches adopt
the same covariance estimate step, but differ in the ensemble update step. Regardless of the
specific approach employed, the goal is to obtain a Bayesian estimate of the
state as efficiently as possible. In many real-world examples these two
approaches perform quite similarly. A more detailed description EnKF may be
found in the book by \cite{Evensen-2009-DAE}.

The EnKF analysis update equations are analogous to the classical KF
equations, except that they use the covariance of the forecast ensemble to
substitute for the matrix $Q$, which in a high-dimensional system is too large
to store. Let $X$ be a random ensemble matrix of dimension $p\times N$ whose
columns are realizations sampled from the prior distribution of the system
state of dimension $p$ with ensemble size $N$. Then the EnKF update formula
is:%
\[
X^{a}=X^{f}+K_{e}(Y-HX^{f}),
\]
where $Y$ is the observed ensemble data matrix whose columns are the true
state perturbed by random Gaussian error. $H$ is, as before, the linear
operator that maps the state vector onto the observational space. The
deviation $Y-HX^{f}$ is commonly referred to as the \textquotedblleft
observed-minus-forecast residual\textquotedblright\ or simply as the
innovation. In the above equation $K_{e}$ is the ensemble Kalman gain matrix
given by%
\[
K_{e}=Q^{f}H^{T}(HQ^{f}H^{T}+R)^{-1},
\]
where $Q^{f}$ is the forecast-error covariance matrix of dimension $p\times
p$, and $R$ is the symmetric and positive-definite observational (measurement)
error covariance matrix. The EnKF technique contains two sources of
randomness: the random model input, and the measurement errors. Assuming that
these two sources of randomness are uncorrelated, the analysis-error
covariance matrix of dimension $p\times N$ can be computed from the equation%
\[
Q^{a}=(I-K_{e}H)Q^{f}.
\]

\subsection{Optimal Statistical Interpolation (OSI)\label{osi}}

In the EnKF, the model error covariance matrix is evolved fully at each data
assimilation step using an MCMC method. In contrast, Optimal Statistical
Interpolation (OSI) is a data assimilation technique based on statistical
estimation theory in which the model error covariance matrix is pre-determined
empirically and is assumed to be time-invariant. The model error covariance
matrix is dependent only on the distance between spatial grid cells. The
correlation length is ad hoc and adjusted empirically.

OSI was derived by \cite{Eliassen-1954-PRC}. This method has
been referred to as \textquotedblleft Statistical
Interpolation\textquotedblright, \textquotedblleft Optimal
Interpolation\textquotedblright, or \textquotedblleft Objective
Analysis.\textquotedblright\ OSI is called univariate if the observations are
of a single scalar field, and multivariate if the
observations of one or more scalar fields are used for estimating another
scalar field \citep{Talagrand-2003-BEO}. Multivariate OSI was developed independently by 
\cite{Gandin-1965-OAM} for the analysis of meteorological fields in
the former Soviet Union. It requires the specification of the cross-covariance
matrix between the observed scalar fields and the scalar field to be estimated%
.

The ensemble OSI (EnOSI), used here, requires much less computational effort than the
EnKF, because the model error covariance matrix is fixed. The EnOSI approach
may provide a practical and cost-effective alternative to the EnKF for
tracking epidemics. The stationary model error covariance matrix in our
epidemic simulation used a version with the correlation function having an
exponential decay along the off-diagonal entries. The ensemble Kalman gain
matrix was then calculated using this time-invariant covariance matrix with a
fixed structure. The accuracy of the EnOSI process will be affected if the
approximate covariance matrix differs substantially from the true covariance
matrix. Therefore, one disadvantage of the OSI is the need for a fixed spatial
covariance structure that can reasonably represent the epidemic dynamics
throughout the whole domain at all times.

The EnOSI analysis update equation, using the stationary covariance, is given
by%
\begin{equation}
X^{a}=X^{f}+K_{OSI}(Y-HX^{f}) \label{eq:OSI}%
\end{equation}
where%
\[
K_{OSI}=Q_{OSI}^{f}H^{T}(HQ_{OSI}^{f}H^{T}+R)^{-1}.
\]

\subsection{Example: An Epidemic Wave Originating In New Mexico}

To test the performance of EnOSI and other tracking algorithms designed for
high-dimensional state vectors, we constructed a spatial simulation of an
epidemic that originates in Santa Fe, New Mexico, and spreads outwards towards
Albuquerque and Denver, Colorado. In this simulation the epidemic moves
smoothly towards Albuquerque, but jumps suddenly to to Denver as if carried by
a traveler in an automobile or airplane. Properly detecting and assimilating
a new feature far from an existing focus is a
serious challenge for EnKF algorithms \citep{Beezley-2008-MEK,Mandel-2010-DDC}.

To improve the realism of the test for the case in which an entirely new
disease emerges for the first time, we initialized all members of the tracking
ensemble so that they contain no disease whatsoever. New data in the form of
an empirical scalar field arrives at time steps 10, 20, 30, 40, and 50. These
data are complete in the sense that in this case $H$ is just the identity
matrix, but they are substantially modified by independent Poisson-distributed
random errors. The tracking algorithm forecasts the state up until the time
when data is received, and then it assimilates this data into the forecast.

\section{Results\label{results}}

We have applied the EnOSI for the New Mexico example mentioned above to the
epidemic model described in Section 2 with an ensemble of size 25. For this
example the \textit{Infected } state of the model is the output of the
observation function. Synthetic data were simulated from a model and
initialized in exactly the same way as the ensemble.

In our epidemics application, the perturbed observations $Y$ in (\ref{eq:OSI})
were obtained by sampling from the Poisson distribution with the intensity
equal to the data value and rounding to integer, consistently with the
stochastic character of the model, instead of Gaussian perturbations as in the
EnKF. This is the key feature behind the successful use of our method and it
also guarantees that $Y$ has nonnegative entries and thus the columns of $Y$
are meaningful as the \textit{Infected} variable. In general, one may have to
guarantee that the entries of the members of the analysis ensemble $X^{a}$ are
also nonnegative, e.g. by censoring. This, however, was not needed in the
results reported here.

The result for each member of the ensemble advanced in time by 10 model time
units is a Bayesian update of the forecast scalar field, which is referred to
as the analysis (i.e. the posterior estimate). We assume that data arrive only
once every 10 time steps, with errors. A total of 5 assimilation cycles were
performed in this manner. The mean and standard deviation (not reported here)
of the ensemble analysis values in each cell of the scalar field gives the
EnOSI estimate of the state of the epidemic, with its uncertainty quantified.
The following figures present a spatial ``image'' of the number of infected
persons over the planar domain considered in the New Mexico example.

%
\begin{figure}
[ptb]
\begin{center}
\fbox{\includegraphics[
natheight=4.753900in,
natwidth=7.199600in,
height=2.879in,
width=4.3474in
]%
{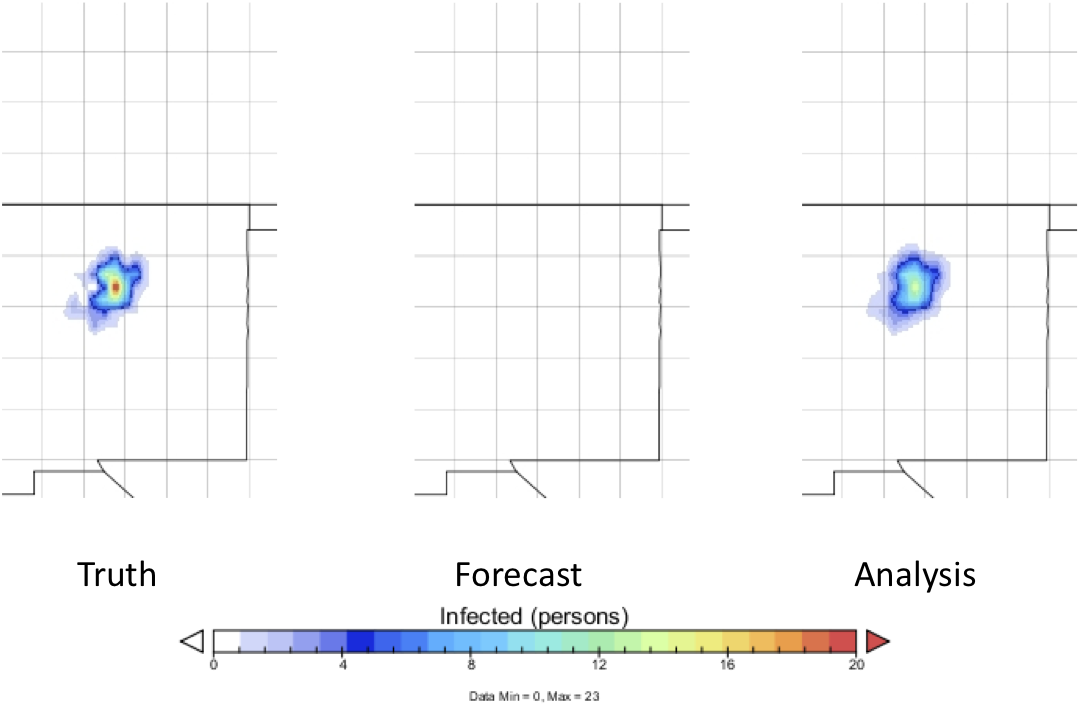}%
}\caption{The epidemic at time 10. (a): The true state of the epidemic.(b):
The \textit{forecast} state of the epidemic prior to arrival of the first
data. (c): The \textit{analysis} state of the epidemic, after data
assimilation.}%
\end{center}
\end{figure}

\begin{figure}
[ptb]
\begin{center}
\fbox{\includegraphics[
natheight=4.753900in,
natwidth=7.199600in,
height=2.879in,
width=4.3474in
]%
{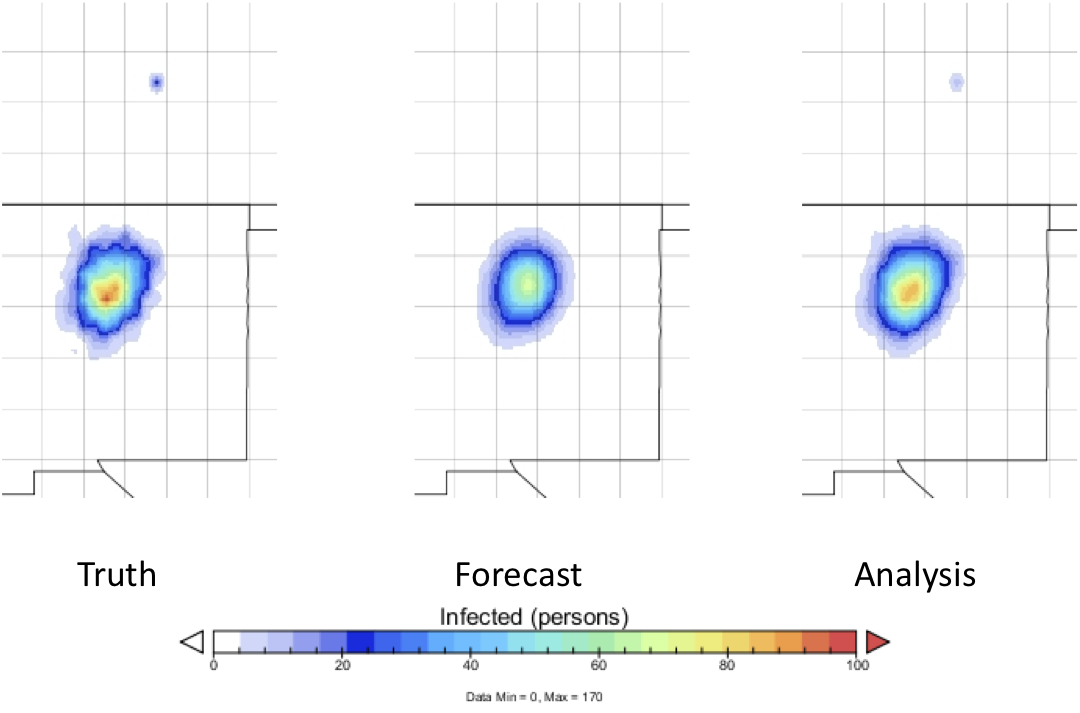}%
}\caption{The epidemic at time 20. A new focus of infection has appeared in
Denver, and is successfully reflected in the analysis state.}%
\end{center}
\end{figure}

\begin{figure}
[ptb]
\begin{center}
\fbox{\includegraphics[
natheight=4.753900in,
natwidth=7.199600in,
height=2.879in,
width=4.3474in
]%
{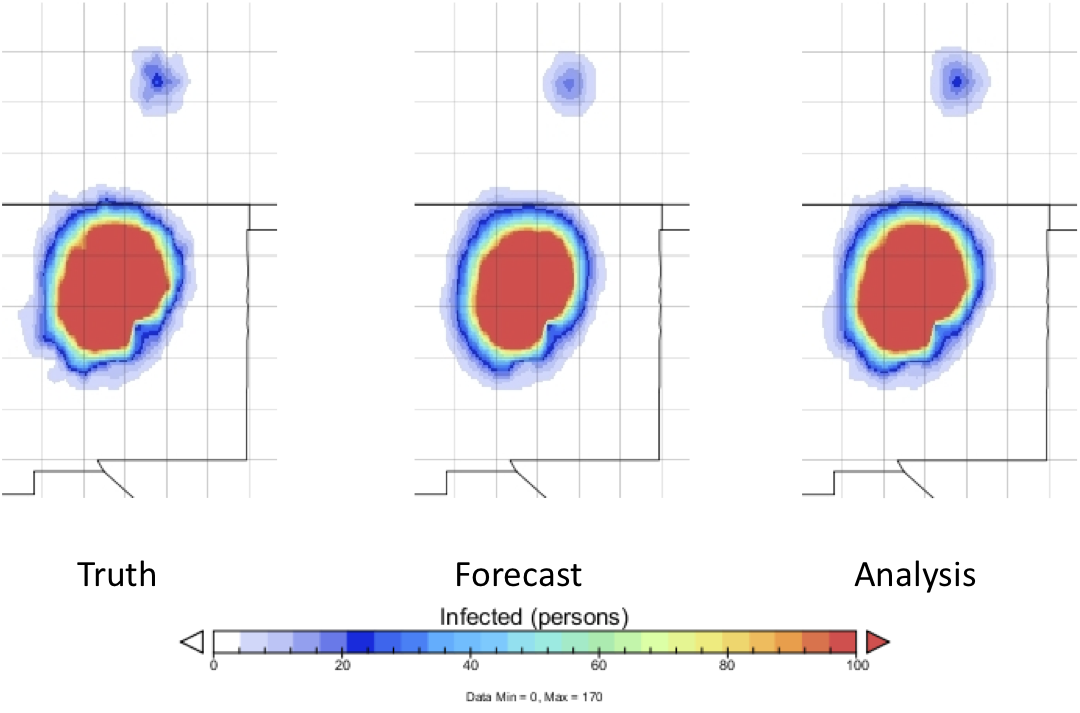}%
}\caption{The epidemic at time 30. The infected zone is expanding rapidly.}%
\end{center}
\end{figure}

\begin{figure}
[ptb]
\begin{center}
\fbox{\includegraphics[
natheight=4.753900in,
natwidth=7.199600in,
height=2.8712in,
width=4.3474in
]%
{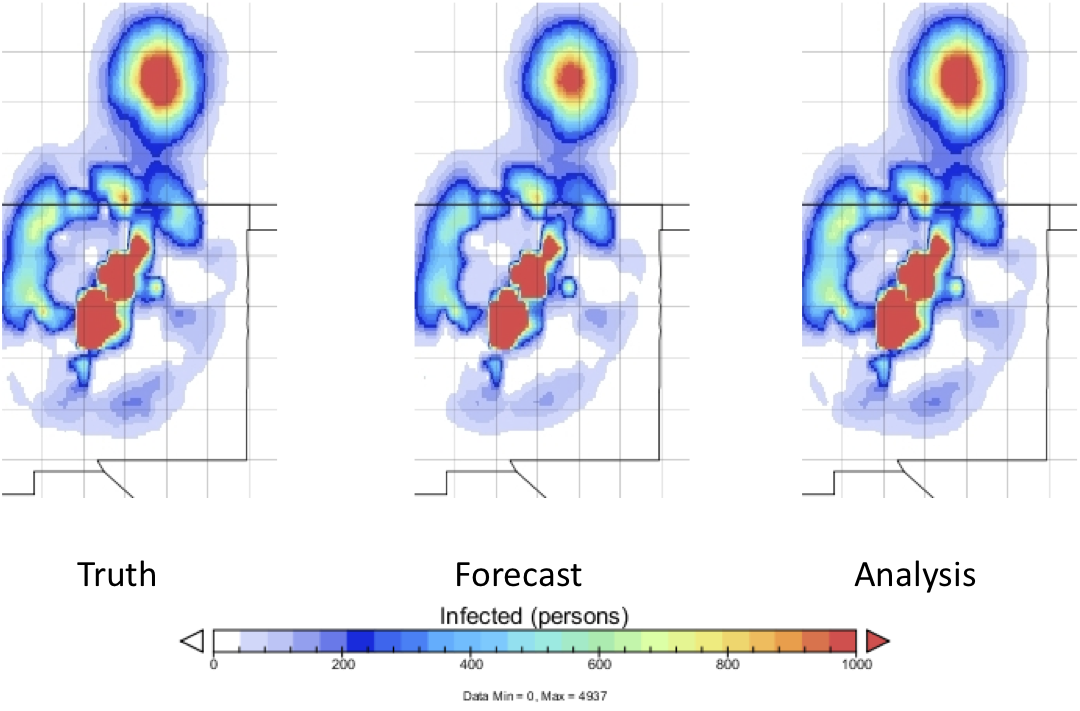}%
}\caption{The epidemic at time 50. Both the forecast and analysis estimates of
state now accurately reflect the true state of the epidemic.}%
\end{center}
\end{figure}
%
Figure 2a shows the epidemic position in Santa Fe, New Mexico, at time 10. The
initial forecast (2b) is empty, as it should be for the first appearance of
any newly emergent disease. The EnOSI analysis (2c) shows that the arriving
data have been partially assimilated, with a resulting picture that is
indistinct and less than fully accurate.
Figure 3a (time 20) shows the growth of the epidemic towards Albuquerque, and
a very small new focus of infection in Denver. The forecast handles the
movement towards Albuquerque quite well, but is devoid of any infection in
Denver. After assimilation of the data, the analysis now also reflects a small
focus in Denver.
Figure 4 (time 30) shows the epidemic gaining size, and beginning a major
expansion within both Albuquerque and Denver.
Figure 5 (time 50) shows the epidemic gaining definition within the most
heavily populated urban regions. The analysis steps, after data assimilation,
are now tracking the epidemic quite~well.

\section{Conclusion\label{conclusion}}

The spread of newly emerging infectious diseases pose a serious challenge to
public health in every country of the world. Tracking the spread of an
epidemic in real-time can help public health officials to plan their emergency
response and health care. The purpose of this paper has been to present the
some preliminary results on the statistical tracking of emerging epidemics of
infectious diseases using a Bayesian data assimilation technique called
ensemble optimal statistical interpolation (EnOSI). Our simulation results
confirms that EnOSI can be used to track the spatio-temporal patterns of
emerging epidemics. We found that EnOSI can efficiently adjust its estimated
spatial distribution of the number of infected, if and when the epidemic jumps
from city to city, and with data that are sparse and error-ridden. The
tracking accuracy in our simulations provides evidence of the good performance
of the EnOSI approach, therefore the assumption that the model error
covariance is time-invariant is reasonable. However, the effect of this
assumption needs more rigorous theoretical justification.

The EnOSI as presented here requires the manipulation of the state covariance
matrix, which gets very demanding if stored as a full matrix - computational
grid of 200 by 200 points results in 40,000 variables and thus 40,000 by
40,000 covariance matrix, which requires a supercomputing cluster.
Sparsification of the covariance matrix can decrease the computational cost
somewhat, but it is still significant and the implementation grows
complicated. For this reason, we plan to investigate a version of OSI by \cite{Mandel-2010-O} with the
covariance implemented by the Fast Fourier Transform (FFT), similarly as in
the FFT EnKF \cite{Mandel-2010-DDC,Mandel-2010-DAM,Mandel-2010-FFT}.

Our research has set the groundwork for further efforts to incorporate the
ideas of data assimilation to track diseases in real-time. Our future work
includes extending the R framework that we have developed for employing an
ensemble of spatial simulations to track diseases to test and compare a other
variants of the EnKF
\citep{Anderson-2001-EAK,Tippett-2003-ESR,Beezley-2008-MEK,Mandel-2009-DAW,Ott-2004-LEK,Hunt-2007-EDA}%
. These variants aim to enhance the performance of the ensemble filters by
representing the underlying model error statistics in an efficient manner.
However, since the ensemble size required can be large (easily hundreds)
\cite{Evensen-2009-DAE} for the approximation to converge \citep{Mandel-2009-CEK}, the amount of computations in the ensemble-based
methods can be significant, and so special localization techniques, such as
tapering, need to be employed to suppress spurious long-range correlations in
the ensemble covariance matrix \citep{Furrer-2007-EHP}. Thus the EnKF (and its
variants) may not work well for problems with sharp coherent features, such as
the traveling waves found in some epidemics. Choosing a small ensemble size,
so small that it is not statistically representative of the state of a system,
leads to underestimation of the analysis error covariances. Choosing a really
large ensemble size may not be computationally feasible and cost efficient.
Finally, methods for incorporating long-distance human
movements to track the rapid geographical spread of infectious diseases have
been proposed in the literature \citep{Brockmann-2009-HMS, Belik-2009-IHM,
Merler-2010-RPH, Balcan-2010-MSS, Belik-2010-HMS}. In the future, we plan to
explore such spatially extended epidemic models to track emerging epidemics.

\bibliographystyle{asa}
\bibliography{epienkf}

\end{document}